\documentclass[prd, showpacs, twocolumn, amsmath,amssymb, widetext, epsfig, nofootinbib, floatfix]{revtex4}
\usepackage{amsmath} 
\usepackage{amssymb} 
\usepackage{graphicx}
\usepackage{dcolumn}
\usepackage{bm}
\usepackage{rotating}
\usepackage[dvipdf]{epsfig}
\usepackage{color}
\usepackage{subfigure}

\newcommand{\potential}{{A}}
\newcommand{\shift}{{B}}
\newcommand{\curvature}{{H}_L}
\newcommand{\shear}{{H}_T}
\newcommand{\kh}{k_H}
\newcommand{\ck}{c_K}
\newcommand{\tppi}{\ck p_T \Pi_T}

\begin{document}


\title{\boldmath Exploring a new interaction  between dark matter and dark energy using the  growth  rate of structure}

\author{Mart\'in G. Richarte}\email{martin@df.uba.ar}
\affiliation{Departamento de F\'isica, Facultad de Ciencias Exactas y Naturales,
Universidad de Buenos Aires and IFIBA, CONICET,Ciudad Universitaria 1428, Pabell\'on I,  Buenos Aires, Argentina}

\author{Lixin Xu}\email{lxxu@dlut.edu.cn (corresponding-author) } %
\affiliation{Institute of Theoretical Physics, School of Physics and Optoelectronic Technology, Dalian University of Technology, Dalian, 116024, People's Republic of China}
\affiliation{State Key Laboratory of Theoretical Physics, Institute of Theoretical Physics, Chinese Academy of Sciences, Beijing 100190, People's Republic of China}

\date{\today}

\begin{abstract}
We present a  phenomenological interaction with  a scale factor power law form  which leads to  the appearance of two kinds of perturbed terms, a scale factor spatial variation  along with   perturbed Hubble expansion rate.   We study both the background and the perturbation evolution within the parametrized post-Friedmann scheme, obtaining that  the exchange of energy-momentum can flow from dark energy to dark matter in order to  keep dark energy and dark matter densities well defined at all times. We  combine several measures of the  cosmic microwave background (WMAP9+Planck) data, baryon acoustic oscillation measurements, redshift-space distortion data,  JLA sample of supernovae, and Hubble constant  for constraining the  coupling constant and  the exponent provided both parametrized the interaction itself. The joint analysis of  ${\rm Planck+WMAP9+BAO}$  ${\rm +RSD+JLA+HST}$ data seems to favor large interacting coupling, $\xi_c  =  0.34403427_{-    0.18907353}^{+    0.14430125}$ at 1 $\sigma$ level, and prefers  a power law interaction with a negative exponent,  thus $\beta=   -0.50863232_{-    0.40923857}^{+    0.48424166}$ at 1 $\sigma$ level.  The CMB temperature power spectrum indicates that a large  coupling constant produces a shift of  the acoustic peaks and affects their amplitudes at lower multipoles. In addition, a larger  $\beta$ exponent generates   a  shift of  the acoustic peaks, pointing a clear deviation with respect to  the  concordance model.  The matter power spectrum are sensitive to the variation of the coupling constant and the $\beta$ exponent. In this context, the interaction alters the scale of matter and radiation equality and  pushes  it away  from the present era, which in turn generates a shift  of the turnover point toward to smaller scale.
\end{abstract}

\pacs{draft}
\bibliographystyle{plain}
\maketitle

\section{Introduction}

Our current view of the Universe is based on the large amount of cosmological observations provided by several surveys \cite{ref:obs}, \cite{ref:wmap9},      \cite{ref:sne3}.  In particular, the first released data by Planck mission seem to indicate that our Universe is filled with two dark components representing the 95  $\%$   of the cosmic budget \cite{ref:planck1}, \cite{ref:planck2}, \cite{ref:planck3}, in fact,  these components  are non-baryonic in nature. On one side,  dark energy can be treated as a fluid with negative pressure  which violates the strong energy condition, counter-balancing the net  gravity effect on the large scale by driven the Universe toward an accelerated expansion era,  leaving behind an earlier non-accelerated epoch. However, it was not discovered yet  what is the microscopic mechanism which triggers the expansion of the Universe and therefore there is not a single fundamental theory which  shows how  fast this transition takes place or when exactly occurs. On the other side, dark matter can be accommodated as a pressureless fluid  which helps to cluster matter  and  affect in this way the whole distribution of galaxies in the Universe  \cite{ref:galaxy4}, \cite{ref:galaxy4b}, \cite{ref:galaxy4c}.  Such component leaves its imprint in the power spectrum of clustered matter  or through gravitational lensing of mass distribution \cite{ref:obs}, \cite{ref:planck2}. A further exploration of  the clustered matter dynamic  needs to  include observations that account for the behavior of  dark matter perturbation and its clustering effect on sub-Hubble scale. A useful manner to achieve this goal is by redshift space distortion (RSD) measurements  of $f\sigma_8$ quantity  at different redshifts, where $f$ is the growth rate factor and $\sigma_8$ is rms density contrast  within an 8 $\rm {Mpc h}^{-1}$ sphere, while $\rm h$ stands for the dimensionless Hubble parameter \cite{ref:distor1}, \cite{ref:distor2}, \cite{ref:distor3}.  Given  RSD measurements are related with peculiar velocities of galaxies, the use $f\sigma_8$ data   is considered as a promising way to  test a cosmological model  beyond the background  level,  helping to break the degeneracy between  different  parameters  \cite{ref:dis1}, \cite{ref:dis2}, \cite{ref:dis3}, \cite{ref:dis4}.

A pressing issue of modern cosmology relies on the existence of non vanishing interaction between dark matter and dark energy. This phenomenon  is  not only  consistent with recent Planck cosmological data  \cite{ref:dis5}, \cite{ref:dis6}, \cite{ref:dis7}, but also seems to be possible at theoretical level  when coupled scalar fields are considered \cite{ref:A0}, \cite{ref:CH1},   \cite{ref:ko1}, \cite{ref:skor1}.  A signature of the exchange of energy in the dark sector can be searched in different manners, for instance,   a starting point  is to explore   the behavior of  dark energy density  at early time \cite{ref:edeC} along with   the amount of dark energy  at the recombination epoch \cite{ref:planck2}, \cite{ref:edeA1}, \cite{ref:edeA2}. Another interesting route  is to examine the impact  of the cosmological constraints over an interacting dark energy model when  a dynamical probe such as  RSD measurements are included \cite{ref:dis7}. As an exmple,  a  geometric (background) test based on $\rm {CMB+BAO+SNe}$ data for an interaction proportional to dark matter density seems to avoid  large coupling, leading to a coupling constant $\xi_c=−0.0013 \pm  0.0008$ at $1\sigma$ level \cite{ref:xinz}.  The inclusion of  RSD measurements  improves a little bit this constraint  provided the joint  statistical analysis performed with $\rm {CMB+BAO+SNe+RSD}$ data leads to $\xi_{c}=0.00140_{-0.00080}^{+0.00079}$ at $1\sigma$ level \cite{ref:me}. Even though the latter result  differs from the former one,  we notice that  the disagreement is not bigger than $0.022\%$. In the light of previous outcome, one might wonder if the small value of the coupling constant  can be  extrapolated as a general output of interacting dark energy models when  dark matter and dark energy are both treated as fluids.  Contrary to the previous results, Salvatelli \emph{et al.} found that a piecewise vacuum interaction, with a large negative coupling constant, is favored   at late time by the joint analysis of   ${\rm Planck+Union2.1+RSD+BAO}$ data \cite{ref:wands}. The latter result would show that  the coupling constant strenght  depends on several factors, that is, it is  likely  linked with the specific form of the interaction, the data selected for implementing a statistical analysis, and above all it is not only  connected with the underlying framework used for treating the dark components \cite{ref:me}.

In this work, we are going to  explore  an alternative interaction between dark matter and dark energy which is  parametrized as a product of the Hubble function and  a scale factor power law up to a coupling constant.   We will include linear perturbations into the cosmic constraint with the help of  the  parametrized post-Friedmann (PPF) formalism \cite{ref:oppf1}, \cite{ref:oppf2}, \cite{ref:sko2} for avoiding  non-adiabatic instabilities which could  happen within the framework of an interacting dark sector \cite{ref:xinz}, \cite{ref:ins4a}, \cite{ref:ins4b}, \cite{ref:ins4c}.    We are going to assess how the interaction affects the  background evolution along with  its impact on  the perturbation equations in the IPPF scheme \cite{ref:xinz}, \cite{ref:me}.  We will analyze how the cosmic parameters are affected by this exchange of energy    taking into account that the perturbed fluid equations will admit a perturbed interaction which  have both  spatial variation of the scalar factor along with perturbed Hubble expansion rate.  We will consider a momentum transfer potential corresponding to an interaction vector proportional  to dark matter velocity in order to avoid a violation of the weak equivalence principle. We will perform a MCMC statistical analysis by modifying CosmoMC package \cite{ref:code2}. Our analysis  will rely on  different observational probes such supernovae Ia from JLA sample, BAO, HST, Planck+WMAP plus a dynamical test based on RSD measurements. We also will explore the TT power spectrum of cosmic microwave radiation and the power spectrum of clustered matter by running a modified version of the CAMB code \cite{ref:code1}, focusing on the behavior of both magnitudes when the coupling constant  varies.


\section{Background and Perturbation Equations}
\subsection{Background}
We assume a  spatially flat  Friedmann-Robertson-Walker background described by the  metric $g_{\mu\nu}=a^{2}{\rm diag}[-1, \gamma_{ij} ]$ with local coordinates $(\eta, x^{i})$, where the  conformal time is defined  as $d\eta =  a^{-1}dt$, and  $\gamma_{ij}$ stands for the 
the spatial metric. Einstein equations, $G_{\mu\nu}=\kappa T_{\mu\nu}$ with $\kappa=8\pi G$, lead to  Friedmann constraint: 
\begin{equation}
H^2 + {K \over a^{2}} = {\kappa \over 3} (\rho_T+\rho_x)\,.
\label{eqn:A}
\end{equation}
As is customary, the continuity equations for  total matter (dark matter, baryons, neutrinos) and effective dark energy are given by  
\begin{align}
\rho_T' &= -3(\rho_T+p_T) + {Q_{c}\over H}\,, \nonumber\\
\rho_x' &= -3(\rho_x+p_x) + {Q_{x}\over H}\,,
\end{align}
when   dark matter and  effective dark energy are coupled. For simplicity, we  assume a phenomenological interaction that can be parametrized in terms of two constants only and that the form of the coupling between  dark components is of a power law form for concreteness:   $Q_{x}= H\bar{Q}_{x}$  with $\bar{Q}_{x}=-\xi_{c}\rho_{\rm c0}a^{\beta}$.  We solve background equations for the interacting dark sector \cite{ref:chimento1}:
\begin{align}
\rho_x&= \rho_{x0}a^{-3(1+w_x)}-{\xi_{c} \rho_{c0} \over  3(1+w_{x})+\beta }a^{\beta}\,,\nonumber\\
\rho_c&= \rho_{c0}a^{-3(1+w_c)} + {\xi_{c} \rho_{c0} \over  3(1+w_{c})+\beta }a^{\beta}.
\label{eqn:beq}
\end{align}
As an illustration, we write the cold dark matter density as the product of its density by its mass, $\rho_c=n_cm_c(a)$, and we realize that the mass of cold dark matter particle is no longer  constant  but  it  varies with the cosmic time:
\begin{align}
m_c (a)&= \rho_{c0}\Big(1  + {\xi_{c}\over  3\beta }a^{\beta+3}\Big).
\label{eqn:MassDV}
\end{align}
At this point we would like to emphasize that dark matter and dark energy are both treated as fluids because the true nature of these dark components is not known yet provided the degree of freedoms which describe them have not been discovered so far. Furthermore, an interaction like we have proposed might not have a simple identification in terms of a coupled quintessence model \cite{ref:A0}, \cite{ref:CH1},  \cite{ref:ko1}, \cite{ref:skor1}. Now,  we focus  on the possible choices that the parameter space would admit and make  some comments about these cases. In the case of  $w_c=0$ and $\xi_{c}>0$,  the parameter space  is  defined through the conditions  $\beta>-3$ and $w_x< -(1+\beta/3)<0$, which allow us to have positive energy density at all times.  Another possible branch corresponds to taking  $w_c=0$ and $\xi_{c}<0$ in the background equations (\ref{eqn:beq}), however, this  case leads to an unphysical conditions $w_x>0$.  In short,  the interaction proposed here only allows to have  an exchange of energy-momentum from dark energy to dark matter ($\xi_{c}>0$) in order to keep the energy densities as positive definite quantities for all times. Besides, the vanilla model can be obtained when the coupling constant vanishes together with the choices $w_c=0$ and  $w_x=-1$. 
\subsection{Perturbation}

As we are interested in analyzing the behavior of  scalar linear  perturbation,  we have to perturb the  Einstein equation and  balance equations around  a FRW background. To do that,  we conveniently write the perturbed variables as  a linear combination of the    eigenfunction (plane-wave) $Y(\bold{x})$ of the Laplace operator \cite{ref:pert1}, \cite{ref:pert2}.  At the end,  the perturbed metric  is characterized by  four functions called potential $A$, shift $B$, curvature $H_{L}$, and shear $H_{T}$: 
\begin{align}
\delta {g_{00}} &= -a^{2} (2 {\potential}Y)\,,~\delta {g_{0i}} = -a^{2} {\shift} Y_i\,,\nonumber\\
\delta {g_{ij}} &= a^{2} (
        2 {\curvature} Y \gamma_{ij} + 2 {\shear Y_{ij}})\,.
\label{eqn:metric}
\end{align}
Here  $Y_{i}$ stands for  the covariant derivative of the eigenfunction $Y(\bold{x})$. The perturbed energy-momentum tensor involves isotropic  and anisotropic pressure perturbations, density, and velocity perturbations:
\begin{align}
\delta{T^0_{\hphantom{0}0}} &=  - { \delta\rho}\,,~\delta{T_0^{\hphantom{i}i}} = -(\rho + p){v}Y^i\,, \nonumber\\
\delta {T^i_{\hphantom{i}j}} &= {\delta p}Y  \delta^i_{\hphantom{i}j} 
	+ p{\Pi Y^i_{\hphantom{i}j}}\,.
\label{eqn:dstressenergy}
\end{align}
The Einstein equations can be written as 
\begin{align}
& {H_L}+ {1 \over 3} {H_T}+   {B \over \kh}-  {H_T' \over \kh^2}=\nonumber\\ 
&{ \kappa \over 2H^2 \ck \kh^{2}}   \left[ {\delta \rho} + 3  (\rho+p){{v}-
{B} \over \kh }\right] \,,
\nonumber\\
& {A} - {H_L'} 
- { H_T' \over 3} - {K \over (aH)^2} \left( {B \over k_H} - {H_T' \over k_H^2} \right)\nonumber\\
&=  {\kappa \over 2H^2 } (\rho+p){{v}-{B} \over \kh} \,,
\nonumber\\
\label{eqn:einstein}
\end{align}
where the prime refers to a logarithmic derivative, $\kh = (k/aH)$,  $c_K = 1-3K/k^2$, and $K$ is the spatial curvature.  In order to further illustrate how the perturbation theory works,  we need to  propose a  covariant  four-vector interaction,  $Q^{\nu}_{~I}$, which encodes the transfer of energy and momentum:
\begin{equation}
\nabla_\nu T^{\mu\nu}_{~I}=Q^{\nu}_{~I},~~~~~~~\sum_{I=x,c} Q^{\nu}_{~I}=0 \,.
\end{equation}
Then, we have a perturbed continuity and a perturbed Navier-Stokes equation for each fluid
 \begin{align}  \label{eqn:conservationq2}
& {(\rho_{i}\Delta_i)'}
	+  3({\rho_{i}\Delta_i}+ {\Delta p_i})+(\rho_i+p_i)(\kh {V}_i + 3 H'_L) \nonumber\\
&	={\Delta Q_{i}-\xi Q_{i} \over H}\,, \nonumber \\
&     {[a^4(\rho_i + p_i)({{V_i}-{B}})]' \over a^4\kh}- { \Delta p_i }+{2 \over 3}\ck p_i {\Pi_i} -(\rho_i+ p_i) A\nonumber\\
&= {a \over k}[Q_{i} (V-V_{T})+f_{i}]\,.
\end{align}

For scalar perturbations (\ref{eqn:metric}), the  perturbed  velocity vector becomes   $u_{\nu~I}=a \big[-(1+AY); (V_{I}-B)Y_{i}\big]$, which means the four vector interaction can be  written as
\begin{align}
Q_{\nu~I}=a\big(-Q_{I} (1+YA)-\delta Q_{I}Y;[F_{I}+Q_{I}(V-B)]Y_{i}\big)\,,
\label{eqn:masa}
\end{align}
where $\delta Q_{I}$ and $F_{I}$ stand for the energy transfer perturbation and the intrinsic momentum transfer potential of $I$-fluid, respectively. The conservation of total energy-momentum for dark components implies that 
\[Q_{c}=- Q_{x}, ~~~F_{c}=-F_{x}, ~~~~~~ \delta Q_{c}=- \delta Q_{x}. \]

\subsection{The PPF method}
 To apply the PPF method  we must make some assumptions about the behavior dark energy  perturbations on super-horizon scales, that is, 
the effective dark energy contribution in the large scale limit ($k_{H}\rightarrow 0$) must be  accommodated in terms of  a single function $f_\zeta(a)$  \cite{ref:oppf2}: 
\begin{equation}
\lim_{k_H \ll 1}
 {\kappa \over 2H^2} (\rho_x + p_x) {V_x - V_T \over k_H}
= - {1 \over 3} \ck  f_\zeta(a) k_H V_T\,.
\label{eqn:L12}
\end{equation}
Here we implicitly  assumed  to be working in the co-moving gauge ($B=V_{T}$, $H_{T}=0$) where we  called  $A=\xi$, $H_L=\zeta$. Further,  the derivative of curvature perturbation  becomes 
\begin{align}
\lim_{k_H \ll 1} \zeta'  &= - {{a \over k}[Q_{c} (V-V_{T})+f_{c}] +  { \Delta p_T }- {2 \over 3}\ck p_T {\Pi_T} \over (\rho_T+ p_T)}\nonumber 
\\
&- {K \over k^2} k_H V_T  +{1 \over 3} \ck  f_\zeta k_H V_T \,,
\label{eqn:zetaprimesh}
\end{align}
so the PPF method  deals with   $\zeta' \simeq 0$ in the co-moving gauge, consequently   dark energy fluctuations become important  at  second order in the co-moving wave number $k_{\rm H}$ (cf.\cite{ref:oppf2} ) provided   $ k_H V_T= {\cal O}(k^{2}_H \zeta)$.  From now on, we are going to neglect any contribution from  anisotropic pressure terms for total matter and effective dark energy, $\Pi_{T}=\Pi_x=0$.

The behavior of the gravitational potential is related with the evolution of curvature perturbation by means of   $\Phi=\zeta + V_{T}/k_{H} $. In fact, the potential  must fulfill a Poisson-like equation  in the quasi-static limit: $\Phi(k_{H}\gg 1)=\kappa a^2 \Delta_T \rho_T/ 2k^2\ck  $. The PPF prescription relies on the simple fact that the behavior of curvature perturbations on the two different scales  must be linked by means of  a single degree of freedom encoded in the $\Gamma$ function \cite{ref:oppf2} as follows: 
\begin{equation}
\Phi +\Gamma = {\kappa a^{2}
\over 2 \ck k^2} \Delta_T \rho_T .
\label{eqn:modpoiss}
\end{equation}
In the large-scale limit, we obtain   $\Gamma'=S-\Gamma$, where the source term can be recast as  
\begin{align}
S=&{\kappa a^2\over 2k^2}\Big({3a \over k \ck}[Q_{c} (V-V_{T})+F_{c}]+{1\over H \ck}[\Delta Q_{c}-\xi Q_{c}]\nonumber\\
&+V_T k_H [-f_{\zeta}(\rho_T+ p_T)+(\rho_x+p_x)]\Big)\,.
\label{eqn:S7gn}
\end{align}
In order to link  the behavior of perturbed quantities  at different scales, we must introduce a master equation for  $\Gamma$:
\begin{equation}
(1 + c_\Gamma^2 k_H^2) [\Gamma' + \Gamma + c_\Gamma^2 k_H^2 \Gamma] = S\,,
\label{eqn:gammaeom}
\end{equation}
where   the condition $c_{\Gamma}k={\cal H}$ determines the transition between large scale regime and the quasi-static phase, that is, the PPF method introduces a new parameter called $c_{\Gamma}$. In regard to the initial condition for  $\Gamma$, we take  a vanishing $\Gamma$ at small scale factor provided the source term goes to zero in this limit.
 
So far we have been working with the function $\Gamma$ in order to connect two separated scales, but  we have not said anything about how we deal  with dark energy perturbations in the previous method. The crucial fact of  the PPF method  is that  forces dark energy density perturbations to remain smaller than the dark matter perturbation \cite{ref:oppf2}. In doing so,  one does not  give a closure relation between dark energy  pressure perturbation and dark energy density perturbation because one would like to  avoid the appearance of  large-scale instabilities \cite{ref:xinz}, \cite{ref:me}. As a result,  the perturbed  dark energy density and peculiar  velocity  are both derived once the new dynamical function is known (cf.  \cite{ref:me}, \cite{ref:oppf2}):
\begin{equation}
\rho_{x}\Delta_{x} + 3(\rho_{x}+p_{x}) {V_{x}-V_{T}\over k_{H} }  = 
-{2k^{2}\ck \over \kappa a^{2}} \Gamma \,, 
\label{eqn:ppffluid}
\end{equation}
\begin{align}
 &\kappa a^2 V_{x}(\rho_{x} + p_{x})  =-{ 2a^2{\cal H}k\over F(a)}\Big( [S_{0}-\Gamma -\Gamma']\nonumber
\\ 
&+{\kappa a^2\over 2k^2}f_{\zeta} (\rho_T +p_T)V_T k_{H}\Big)+ \kappa a^2 V_{T}(\rho_{x} + p_{x})  \,,
\label{eqn:veeff2}
\end{align}
where $F(a) = 1 + 3 {\kappa  a^2 \over 2k^2 \ck} (\rho_T + p_T)$. Here we are keeping only the leading terms  in the source  expression for  small comoving wave number $k_H$, and  we are calling this result $S_{0}$  \cite{ref:me}. 

In this study,  we will be concerned with the physical consequences coming from the IPPF method  when an interaction four-vector  can be written as $Q^{\mu}_{x}=H\bar{Q}_{x}u^{\mu}_{c}$  with  $\bar{Q}_{x}=-\xi_{c}\rho_{\rm c0}a^{\beta}$ and  $Q_{x}\equiv aQ^{0}_{x}= H\bar{Q}_{x}$. In this case,  the interaction  is given by 
\begin{align}
Q_{\nu~x}=a \bar{Q}_{x} H \big[-(1+AY); (V_{c}-B)Y_{i}\big]\,.
\label{eqn:cmasa1}
\end{align} 
From (\ref{eqn:masa}) and (\ref{eqn:cmasa1}), we obtain that the perturbed interaction can be written as 
\begin{align}
\delta Q_{x}=\delta H \bar{Q}_{x}(a)+ H  \bar{Q}'_{x}(a)\delta a\,,
\label{eqn:perq}
\end{align} 
while for the momentum transfer potential  we find that  $F_{x}=Q_{x}(V_{c}-V)=-F_{c}$  provided $Q^{\mu}_{c} || u^{\mu}_{c}$, avoiding in this way the violation of weak equivalence principle \cite{ref:me}. Notice that $\delta Q_{x}$ contains two different kinds of terms, namely  $\delta Q_{x}$ receives contributions from the spatial variation  of $H$ and the spatial variation of the scale factor. In order to  carry on and explore the impact of this new interaction on the parameter space, it is essential to  modify or include several routines in the numerical code called CAMB. Given that this code is written in  the synchronous gauge we need to handle all these expressions into this gauge characterized by $A =  B = 0$. Replacing both conditions  into  gauge transformation for metric variables indicates that scalar perturbations can be treated in terms of two functions:  ${\eta_T} \equiv -\frac{1}{3}  H_T - H_{L}$ and $h_{L} = 6H_L$ \cite{ref:me}. From the later fact, with the help of  the metric (\ref{eqn:metric}), we  calculate $\delta H$  and $\delta a$ in the synchronous gauge as
\begin{align}
[\delta a]_{\rm syn}&= -a(\eta_{T}-{\cal H}\sigma)\,,\nonumber\\
[\delta H]_{\rm syn}&=-{1\over a}[\dot{\eta}_{T}+\sigma ({\cal H}^2-\dot{{\cal H}})] ,
\label{eqn:deltas}
\end{align}
where $\sigma=(\dot{h}_{L}+3\dot{\eta}_{T})/2k^2$ and  the overdot refers to conformal time derivative. Combining  Eqs. (\ref {eqn:deltas}) and Eq. (\ref{eqn:perq}) in the synchronous gauge, we have
\begin{align}
\delta Q_{x}=\frac{\bar{Q}_{x}}{a}[  \dot{\eta}_{T} +\beta{\cal H}\eta_{T} + \sigma ( {\cal H}^2-\beta{\cal H}^2- \dot{{\cal H}}) ].    
\label{eqn:perqF}
\end{align} 

To understand the cosmological constraints performed with redshift space distortion measurements  we must give further insights about the dynamical evolution of  dark matter perturbations.  We need to  examine  if the exchange of energy-momentum in the dark  sector may alter the standard Euler and continuity equations for dark matter, leading to a detectable signature about its dynamic in relation with the behavior of  galaxies.  Our starting point is to calculate the evolution equations for the perturbation of cold dark matter in the synchronous gauge \cite{ref:me}: 
\begin{eqnarray}
\dot{\delta}_{c} +\theta_{c} + {\dot{h}_{L}\over 2}&=&{a\over \rho_{c}} [\delta Q_{c}-\delta_{c}Q_{c}],
\label{eqn:cdmg0}
\end{eqnarray}
\begin{eqnarray}
\dot{\theta}_{c}+ \theta_{c}{\cal H}&=& {a\over \rho_{c}} [Q_c (\theta-\theta_{c})-k^{2}F_{c}],
\label{eqn:cdmg}
\end{eqnarray}
where  we used  that the adiabatic sound speed defined as $c^2_{ac}=\dot{p}_c/\dot{\rho}_c=w_c+\dot{w}_c/(\dot{\rho}_c/\rho_c)$ vanishes, $w_{c}=0$, and the physical sound speed in the rest frame, namely $c^2_{sc}=(\delta p_c/\delta\rho_c)_{\rm rest~frame}=0$  for  cold dark matter, and the density contrast is defined as $\delta_c=\rho_c\Delta\rho_c/\rho_c$. For the definition of $\theta_c$ and $\theta$ see our previous article \cite{ref:me}.

After decoupling, the baryons  do not interact with the dark sector so its Euler and continuity equations  within the synchronous gauge are given by 
\begin{eqnarray}
\dot{\delta}_{b} +\theta_{b} + {\dot{h}_{L}\over 2}&=&0, \\
\dot{\theta}_{b}+ \theta_{b}{\cal H}&=&0.
\label{eqn:barion}
\end{eqnarray}

In our case,    we find $[Q_c (\theta-\theta_{c})-k^{2}F_{c}]=0$ provided  the momentum transfer potential is   $-k^{2}F_{c}=Q_{c}(\theta_{c}-\theta)$. This  implies that the Euler equation for dark matter is not altered in relation with the non-interacting case. As a result,  we obtain that the difference between Euler equations (\ref{eqn:cdmg}-\ref{eqn:barion}) leads to $(\dot{\theta}_c -\dot{\theta}_b)=-{\cal H}(\theta_c -\theta_b)$, and therefore we can assure that there is no velocity bias  within this model \cite{ref:Ro2}.  This fact pinpoints that   dark matter velocity is not directly affected by the interaction, which means that dark matter follows geodesic  and there is not violation of the weak equivalence principle \cite{ref:Ro3}. 


Combining Eq. (\ref{eqn:cdmg0}) for $\dot{\delta}_c$ along with  the Einstein equations in the synchronous gauge \cite{ref:pert2},  we arrived at the second-order evolution equation for dark matter perturbation:
\[\ddot{\delta}_c + {\cal H}\dot{\delta}_c  \Big(1+ \frac{aQ_{c}}{\rho_c {\cal H}}\Big)= \]
\begin{eqnarray}
 4\pi G a^2 \Big[\delta_b\rho_b +\delta_c\rho_c \Big(1- \frac{(\rho^{-1}_cQ_{c})^{\dot{}}}{4\pi G a\rho_c} \Big) \Big] + a \Big(\frac{\delta Q_{c}}{\rho_c}\Big)^{\dot{}}.
\label{eqn:DMP}
\end{eqnarray}
The growth of dark matter perturbations are affected  in a number of ways. Firstly, the modification of the  dark matter contrast density  enters through the expansion history encoded in the Hubble expansion and cold dark matter density, which will lead to a faster or slower clustering of matter particles according to the values taken by $\xi_c$ and $\beta$. Secondly, the friction term of Eq. (\ref{eqn:DMP}) is altered by the extra contribution, $aQ_{c}/\rho_c {\cal H}$, implying that the rate growth will depend both on $Q_c$ and the background evolution described by ${\cal H}$. Thirdly, the cold dark matter particle feel and additional force that is mediated by dark energy particle described by an effective  gravitational constant. As cold dark matter particles have a mass which varies with time,  the uncoupled baryons will feel this effect through a changing gravitational potential. It is clear that the extra-term in Eq. (\ref{eqn:DMP}) is related with the fact that $\delta Q_{c}$ cannot be recast as  $\delta\rho_c Q_{c} $  provided our interaction is no proportional to $\rho_c$. At this point, we must highlight the importance of  dark matter perturbations  and baryon perturbations,  both contributions are  essentials  for understanding the formation of the large-scale structure of the Universe. More precisely,  the growth data involve  the linear  perturbation of the  overall  growth factor $\delta_{\rm m}$  in terms of the function $f_m=d{\rm ln~ \delta_{\rm m}} /d{\rm ln~ a}$ along with  the r.m.s density contrast within an sphere of radius $R_8=8 {\rm Mpc}~ {\rm h}^{-1}$, which is  related with the matter power spectrum.  As explained before, the measurements of  redshift space distortions lead to a bias-independent  quantity called $f(z)\sigma_{8}(z)$ \cite{ref:dis3} and is used it  as a powerful tool for improving the cosmological constraints beyond the background probes based on supernovae data.  It is possible to use these data in our cosmic constraint but we must first modify the CAMB code for extracting  the theoretical value  of $f(z)\sigma_{8}(z)$,  and such procedure  involves to re-write different routines. On  one side,  $\sigma_8 (z)$ is given by 
\begin{equation}
 \sigma_8(z)= \left[\frac{1}{2\pi^2}  \int^{\infty}_{0}{dk k^2 W^{2}_{8}(k)P(k,z)} \right]^{1/2},
\end{equation}
where $W_{8}(k)$ is the Fourier transform of the top-hat window function, with an interval defined by $R_8$, and $P(k,z)$ refers to the matter power spectrum. It can be seen that in our model the overall growth matter rate  is
\begin{equation}
 f_m(a)=\frac{\rho_c\delta'_c+ \rho_b\delta'_b}{{\cal H}(\delta_c+ \rho_b)}+ \frac{\xi_c \rho_{c0}a^{\beta}}{\rho_c+ \rho_b}\left(\frac{\delta_c}{\delta_m}-1\right),
 \label{eqn:GRM}
\end{equation}
where the interaction induces an extra term  which plays the role of a relative weight between cold dark matter contrast density and total matter  contrast density. Moreover,  (\ref{eqn:GRM}) shows that the interaction reduces the overall growth matter rate for positive coupling constant, $\xi_{c}>0$. This difference therefore must leave to a detectable signature, which allows differentiation between previous works \cite{ref:me},  \cite{ref:yangf} .


\begin{table}
\begin{center}
\begin{tabular}{ccc}
\hline\hline z & $f\sigma_8(z)$ & survey  \\ \hline
$0.067$ & $0.42\pm0.06$ & ${\rm 6dFGRS~(2012)}$ \\
$0.17$ & $0.51\pm0.06$ & ${\rm 2dFGRS~(2004)}$ \\
$0.22$ & $0.42\pm0.07$ & ${\rm WiggleZ~(2011)}$ \\
$0.25$ & $0.39\pm0.05$ & ${\rm SDSS~LRG~(2011)}$ \\
$0.37$ & $0.43\pm0.04$ & ${\rm SDSS~LRG~(2011)}$ \\
$0.41$ & $0.45\pm0.04$ & ${\rm WiggleZ~(2011)}$\\
$0.57$ & $0.43\pm0.03$ & ${\rm BOSS~CMASS~(2012)}$ \\
$0.60$ & $0.43\pm0.04$ & ${\rm WiggleZ~(2011)}$ \\
$0.78$ & $0.38\pm0.04$ & ${\rm WiggleZ~(2011)}$ \\
$0.80$ & $0.47\pm0.08$ & ${\rm VIPERS~(2013)}$ \\
\hline\hline
\end{tabular}
\caption{Compilation of  $f\sigma_8(z)$  data points obtained from several galaxy surveys using RSD method.}
\label{tab:fsigma8data}
\end{center}
\end{table}

\begin{table}[tbp]
\centering                                                                                                                 
\begin{tabular}{ccc}                                                                                                            
\hline\hline                                                                                                                    
Parameters & Mean with errors & Best fit \\ \hline
$\Omega_b h^2$ & $    0.02215594_{-    0.00026246}^{+    0.00025836}$ & $    0.02222252$\\
$\Omega_c h^2$ & $    0.11864910_{-    0.00211040}^{+    0.00196349}$ & $    0.11845120$\\
$100\theta_{MC}$ & $    1.04143237_{-    0.00057134}^{+    0.00058065}$ & $    1.04152800$\\
$\tau$ & $    0.08910617_{-    0.01346580}^{+    0.01235947}$ & $    0.08779789$\\
$w_x$ & $   -1.12874223_{-    0.06244387}^{+    0.08205012}$ & $   -1.14908200$\\
$\xi_c$ & $    0.34403427_{-    0.18907353}^{+    0.14430125}$ & $    0.32413630$\\
$\beta$ & $   -0.50863232_{-    0.40923857}^{+    0.48424166}$ & $   -1.06008500$\\
${\rm{ln}}(10^{10} A_s)$ & $    3.08464588_{-    0.02425935}^{+    0.02442454}$ & $    3.08247800$\\
$n_s$ & $    0.96269367_{-    0.00622752}^{+    0.00624026}$ & $    0.96447760$\\
$H_0$ & $   63.68287359_{-    2.30201737}^{+    3.34545416}$ & $   65.75439000$\\
$\Omega_x$ & $    0.64884118_{-    0.02299489}^{+    0.04222175}$ & $    0.67314830$\\
$\Omega_m$ & $    0.35115882_{-    0.04222199}^{+    0.02299549}$ & $    0.32685170$\\
$\Omega_m h^2$ & $    0.14145019_{-    0.00201151}^{+    0.00187374}$ & $    0.14131890$\\
$\sigma_8$ & $    0.76985212_{-    0.02119374}^{+    0.02336652}$ & $    0.78026930$\\
${\rm{Age}}/{\rm{Gyr}}$ & $   13.77835552_{-    0.03804099}^{+    0.03828760}$ & $   13.78876000$\\
$z_*$ & $ 1090.05905721_{-    0.43505566}^{+    0.44064201}$ & $ 1089.95200000$\\
$r_*$ & $  144.94953514_{-    0.47690640}^{+    0.47179303}$ & $  144.94740000$\\
$\theta_*$ & $    1.04165139_{-    0.00056187}^{+    0.00057374}$ & $    1.04174300$\\
\hline\hline                                                                                                                    
\end{tabular}                                                                                                                   
\caption{Statistical results from  the global fitting performed with the Planck 2013+WMAP9+JLA+BAO+RSD+HST data.}\label{tab:results}                                                                                                   
\end{table}

\begin{figure*}[tbp]
\begin{center}
\includegraphics[totalheight=9.6in,width=7in, clip] {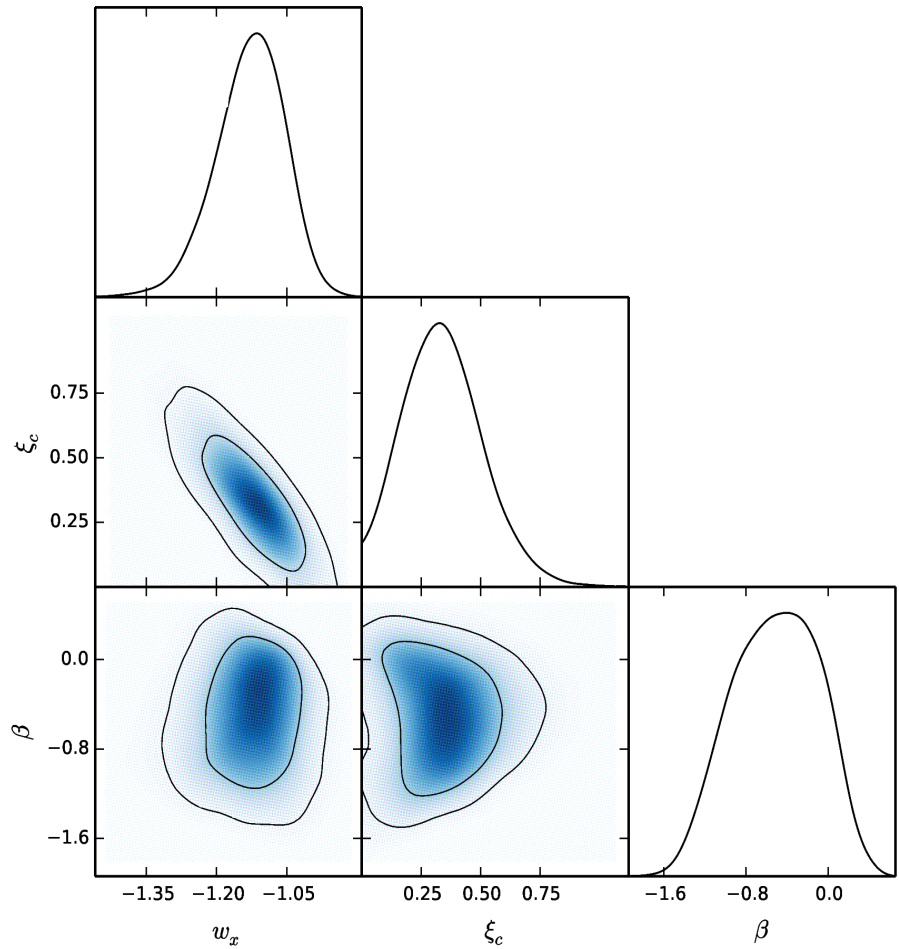}
  \caption{\scriptsize{68 $\%$ and 95 $\%$ constraints on $w_x$, $\xi_c$, and  $\beta$ from the combined analysis made with the Planck 2013+WMAP9+JLA+BAO+RSD+HST data.  The 1D marginalized posterior distribution of $w_x$, $\xi_c$, and  $\beta$ are also shown.}}
\label{fig:pb}
\end{center}
\end{figure*}

\begin{figure*}[tbp]
\centering 
\includegraphics[totalheight=9.6in,width=7in, clip]{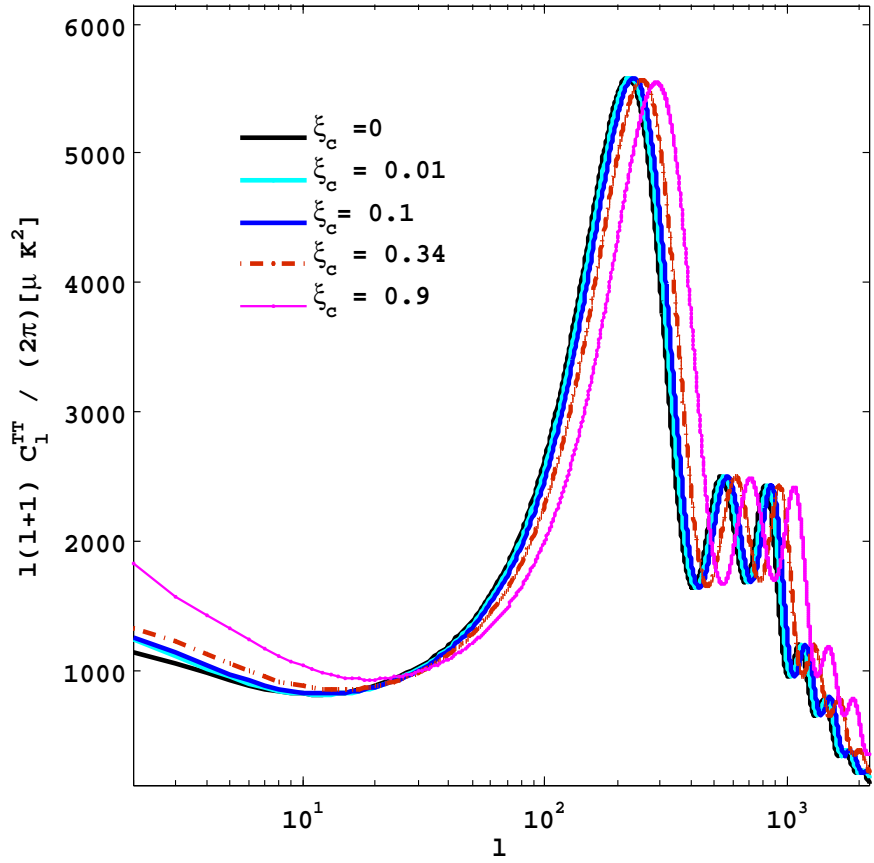}
\caption{\label{fig:c1} Typical effects of the coupling constant on the theoretical CMB temperature power spectrum. We exhibit  $C^{\rm TT}_{\rm l}$ versus multipole for different values of the coupling constant. }
\end{figure*}

\begin{figure*}[tbp]
\centering 
\includegraphics[totalheight=9.6in,width=7in, clip]{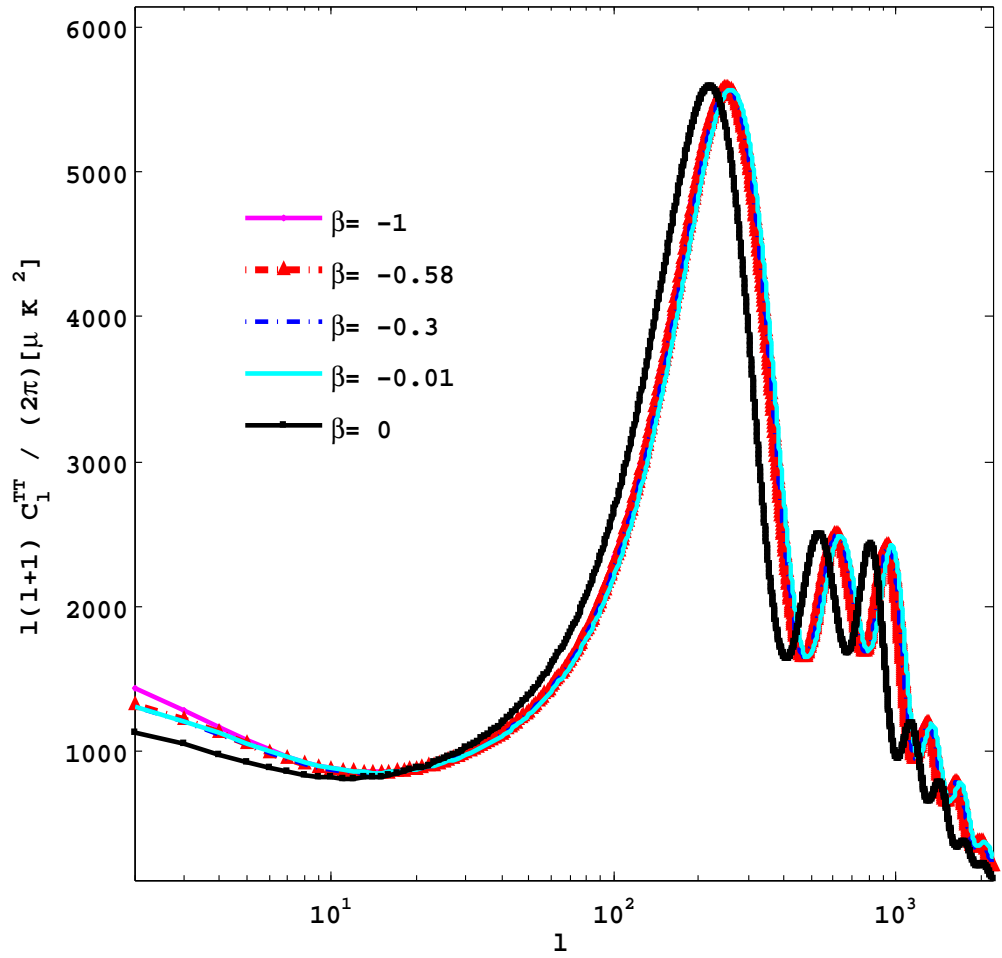}
\caption{\label{fig:c2} Typical effects of the $\beta$ exponent on the theoretical CMB temperature power spectrum. We show $C^{\rm TT}_{\rm l}$ versus multipole for different  power-law interactions.}
\end{figure*}

\begin{figure*}[tbp]
\begin{center}
\includegraphics[totalheight=9.6in,width=7in, clip]{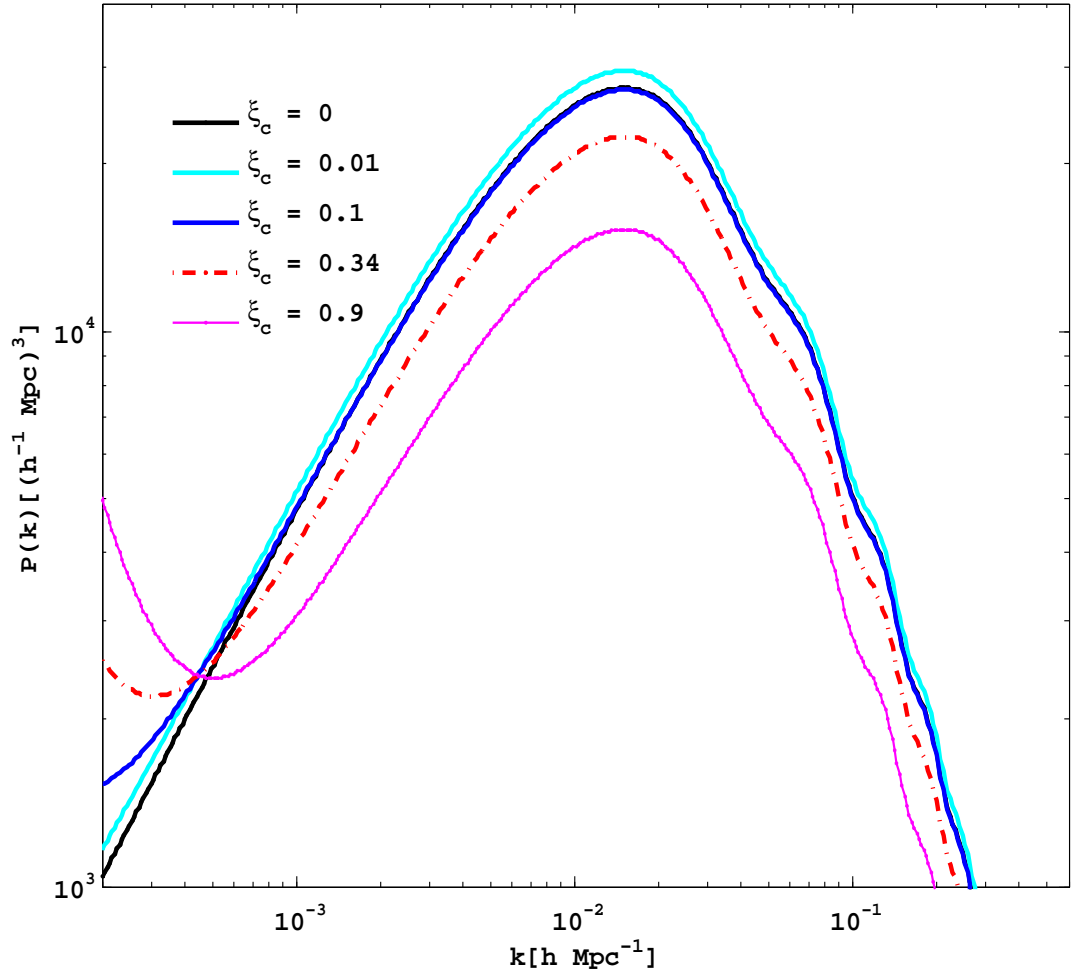}
 \caption{\scriptsize{The total matter  power spectrum at $z=0$ as the coupling constant varies from 0 to 0.9. Baryon acoustic oscillations generate wiggles in the matter power spectrum.}}
\label{fig:p1}
\end{center}
\end{figure*}

\begin{figure*}[tbp]
\begin{center}
\includegraphics[totalheight=9.6in,width=7in, clip]{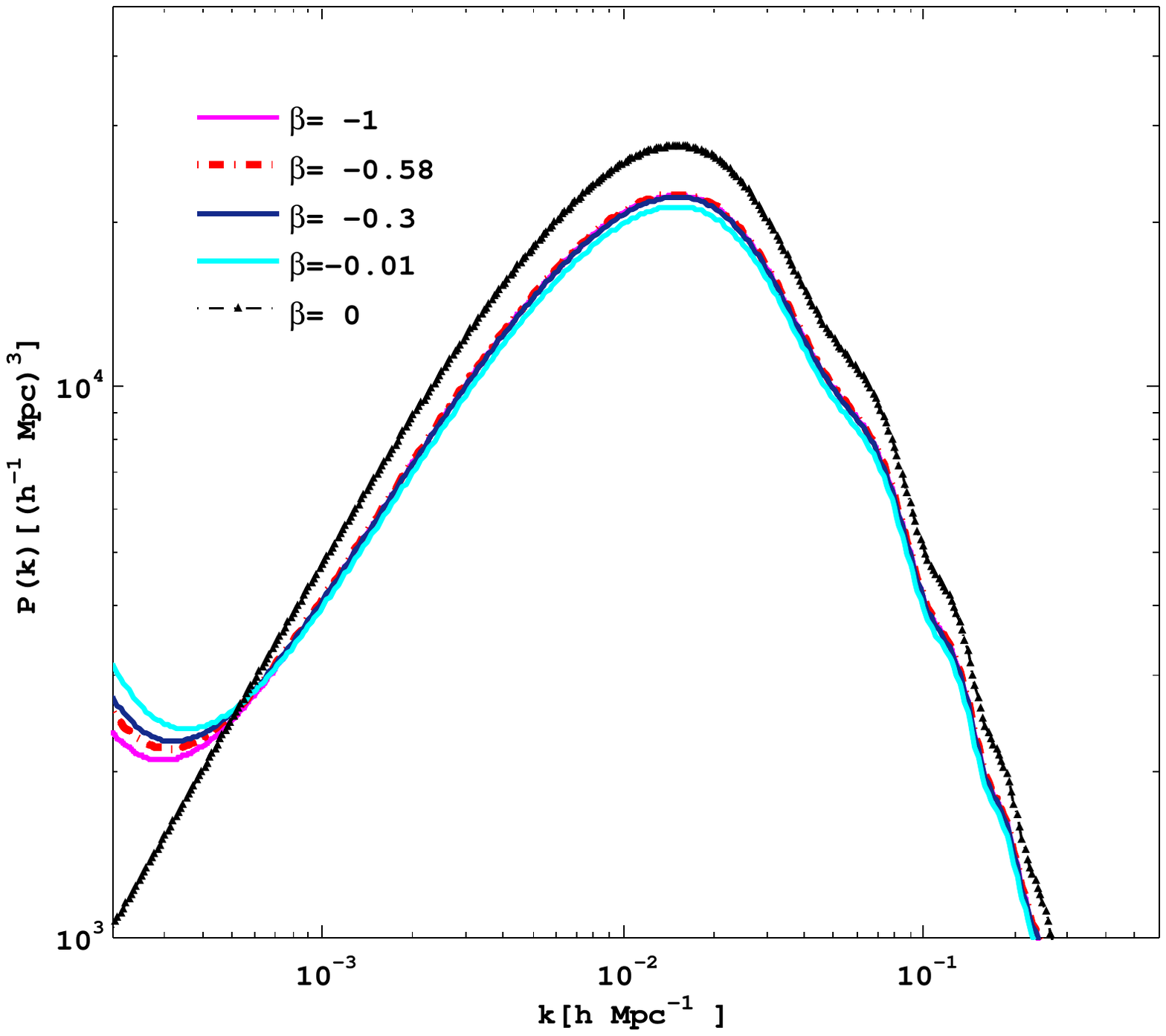}
 \caption{\scriptsize{The total matter  power spectrum at $z=0$ as the $\beta$ exponent  varies from 0 to -1. Baryon acoustic oscillations generate wiggles in the matter power spectrum.}}
\label{fig:p2}
\end{center}
\end{figure*}


\section{Constraint and Results}
 We perform a statistical estimation  of the cosmic parameters by using the Markov Chain Monte-Carlo method with help of  public code CosmoMC \cite{ref:code2} and CAMB  code \cite{ref:code1}. In order to do so,  we add a new module to CosmoMC package for taking into account the value of $f_m$ obtained with  the CAMB code, in this way, we calculate the theoretical value of  $f(z)\sigma_{8}(z)$  at different redshift, compare with their  observational value, and build the  $\chi^2$ likelihood for performing a statistical estimation on the  parameter space \cite{ref:Ya}. Two comments are in order concerning the PPF ``parameters''.  When  $c_{\Gamma}$ is taken as  free model parameter  one finds that its posterior probability distribution has a  flat shape, showing that any value between zero and the unity are physically admissible \cite{ref:me},  in fact,  a similar analysis can be obtained for the function $f_{\xi}$ \cite{ref:oppf2}. The list of the  cosmological  data used for our parameter estimations (cf. \cite{ref:me} for further details about the methodology) are: 
 \begin{itemize}
   \item ${\rm SNe Ia}$, $ {\rm JLA~ data}$: 740 supernovae  samples from low redshift $z=0.02$ to large one  $z \simeq 1$, obtained from the joint analysis of SDSS II and SNLS  \cite{ref:sne3}.
   \item  ${\rm CMB}$, ${\rm WMAP9+Planck}$: multipole measurements obtained by WMAP9 team  \cite{ref:wmap9}  and Planck satellite \cite{ref:planck1}, \cite{ref:planck2}, \cite{ref:planck3}. WMAP9 project involves the measurements of  Atacama Cosmology Telescope (ACT)  at high  multipoles, $\ell \in [500, 10000]$, along with the South Pole Telescope (SPT) observations which reported data over the range   $\ell \in [600, 3000]$. Planck survey  performed measurements over  a complementary zone, $\ell \in [2, 2500]$.
   \item  ${\rm BAO}$, ${\rm DR9-DR7-6dFGS}$: the 6dFGS mission  reported  $d_z(0.106)$ \cite{ref:galaxy4}, SDSS-DR7 measured $d_z(0.35)$ \cite{ref:galaxy4b},  SDSS-DR9 exploration led to $d_z(0.57)$ \cite{ref:galaxy4c},  and diverse measurements from the  Wiggle Z dark energy survey  reported $d_z(0.44)$, $d_z(0.60)$, and $d_z(0.73)$.
   \item  ${\rm HST}$, ${\rm Hubble}$: Gaussian prior for $H_0$.
   \item  ${\rm RSD}$, ${\rm Growth~ data}$: measurements of the quantity $f(z)\sigma_{8}(z)$ at different redshifts unify the cosmic growth rate  $f$ and the matter power spectrum $\sigma_{8}$ normalized with the co-moving scale $8 \rm{h^{-1}~Mpc}$, in a single quantity which includes the latest results of galaxy surveys such as 6dFGS, BOSS, LRG, Wiggle Z, and VIPERS [see  Table (\ref{tab:fsigma8data})].   
 \end{itemize}
The  parameter space  is given by \cite{ref:me}
\begin{align}
{\cal P}=\Big( \Omega_{b}h^{2}, \Omega_{c}h^{2}, 100\theta_{\rm MC}, {\rm n}_{\rm s}, \ln (10^{10}{\rm A}_{\rm s}), \tau,  w_{x}, \xi_{c}, \beta \Big),
\label{eqn:PS}
\end{align}
and  each data base is taken as independent one, so the likelihood distribution  is 
\begin{align}
\chi^2_{\rm {total}}=\chi^2_{\rm {SNe}}+\chi^2_{\rm{BAO}}+\chi^2_{\rm{CMB}}+\chi^2_{\rm{HST}}+\chi^2_{\rm {RSD}}.
\label{eqn:ctotal}
\end{align}
Let us analyze the outcome of our statistical analysis obtained by combining all ${\rm Planck 2013+WMAP9+JLA+BAO+HST+RSD}$ data [see Table (\ref{tab:results})]. The equation of state of dark energy  is 
given by $w_x=   -1.12874223_{-    0.06244387}^{+    0.08205012}$ at $1\sigma$ level, such values confirm that the observational data prefer a phantom-like equation of state  over a non-phantom one [see Fig. (\ref{fig:pb})]. Regarding the amount of dark matter at present time we find that the best fit to all data combined: $\Omega_m= 0.35115882_{- 0.04222199}^{+    0.02299549}$  at $1\sigma$ level, while, the ${\rm Planck+WP}$ data leads to $\Omega_m =   0.315_{- 0.018}^{+0.016}$, then the disagreement  is not bigger than $0.11\%$, moreover, comparing with the Planck 2015 data we obtain a relative difference near $0.22\%$ for the Planck EE+lowP \cite{ref:planck15}. The best fit for dark energy amount is $\Omega_x= 0.64884118_{- 0.02299489}^{+0.04222175}$ at $1\sigma$ level,  showing  a deviation of $0.05\%$ in relation with  the value reported by Planck mission \cite{ref:planck3}. This result is utterly consistent with the interacting cosmology explored here because  the transfer of energy goes from dark energy to dark matter, then  we can expect that such mechanism increases the current value of dark matter. Further, the dynamical probe included in the statistical analysis clearly shows that the overall growth rate of matter (\ref{eqn:GRM}) is quite sensitive to the interaction proportional to the current dark matter amount.  In this regard, the joint analysis of ${\rm Planck 2013+WMAP9+JLA+BAO+HST+RSD}$ data [see Table (\ref{tab:results}) and Fig. (\ref{fig:pb})] shows that a large coupling constant cannot be ruled out in our model provided $\xi_c  =  0.34403427_{-    0.18907353}^{+    0.14430125}$ at 1 $\sigma$ level.  We stress that this result  depends on the addition of growth rate data and the interaction itself. More precisely, if one considers only geometric probes for performing the statistical test with  $\rm {CMB+BAO+SNe}$ data, the coupling constant remains lower, near ${\cal O}(10^{-3})$ \cite{ref:xinz}. Despite  the inclusion of the growth rate data  gives a non zero deviation from the previous result, the strength of the interaction   still remains small, that is,  $\xi_{c}=0.00140_{-0.00080}^{+0.00079}$ at $1\sigma$ level  \cite{ref:me}. Notice that  the inclusion of a perturbed expansion Hubble contribution in the standard perturbation theory does not prefer large interaction even if growth rate data plus geometric probes are taken into account \cite{ref:yangf}. Therefore, the key point for understanding the large value of coupling constant reported here is linked with  some cooperative effects coming from  the inclusion of perturbed expansion Hubble rate along with spatial variation of scale factor within the interacting PPF method. Besides, the combined data of ${\rm Planck 2013+WMAP9+JLA+BAO+HST+RSD}$ prefer a  power law form  with negative exponent, namely  $\beta=   -0.50863232_{-    0.40923857}^{+    0.48424166}$ at 1 $\sigma$ level, indicating that the interaction is stronger in the early universe and weakens at  present time.  However, notice that our finding does not  mean that the interaction has a vanishing  contribution in the latter epoch.    

The impact of the interaction  can be traced by inspecting the changes introduced in the cosmic background radiation and clustered matter power spectrum, thus  we explore how these quantities are affected by the two parameters which describe the interaction itself: $\xi_c$ and $\beta$. Fig. (\ref{fig:c1}) shows the CMB temperature anisotropies in terms of the multipoles for several values of the coupling constant with $\xi_c \in [0, 0.9]$. Firstly, the position of the acoustic peaks shifts to right by increasing the coupling constant. The peaks are located at  $\ell_{n} \simeq n \pi/\theta_{A}$, where the acoustic scale is defined as $\theta_{A}=s(z_{dec})/r_{z_{dec}}$  with $r$ known as the comoving angular diameter distance,   $z_{dec} \simeq 1090$ indicates the epoch when baryons decouple from photons, and the sound horizon is given by 
\begin{align}
s(z)= \int_{0}^{1/(z+1)}{ c_{s}(a)\frac{da}{a^2 H(a)}},
\label{eqn:defs}
\end{align}
where the sound speed for the photon-baryon fluid is $c_{s}=c/\sqrt{3(1+R)}$ and $R=3a\Omega_b/4\Omega_r$. Since the coupling constant is positive, the exchange of energy goes from dark energy to dark matter,  increases linearly the amount of dark matter at early time, affecting  the sound horizon at the end, which modifies the value of the first peak located at $\ell_{1} \simeq  \pi/\theta_{A}$ as it can be seen from Fig. (\ref{fig:c1}) by comparing the red-line  ($\xi_c=0.34$) with the black-line ($\xi_{c}=0$). A secondary effect refers to the amplitude of CMB power spectrum at lower multipoles, it turns out that  increasing the coupling constant shows a deviation from the vanilla model, so we believe that such phenomenon can be properly explored in the future by including the galaxy-ISW cross-correlated power spectrum \cite{ref:dis7}. Besides, Fig. (\ref{fig:c2}) shows a shift to right in the acoustic position peaks  as one moves  from a vanishing exponent to a negative exponent equal to $-1$ while the other parameters are fixed to our best cosmology.  This effect can be understood easily by taking into account that the extra-term in the cold dark matter density increases from a constant value to $a^{-1}$,  then it amplifies the amount of dark matter at early times. It is clear that changes  in the background expansion history of the Universe and modifications to the growth rate of matter perturbations will affect the matter power spectrum. In Fig. (\ref{fig:p1}) we plot the matter power spectrum at $z=0$  generated using CAMB for different values of the coupling constant, while,  Fig. (\ref{fig:p2}) depicts the behavior of $P(k)$ when the exponent $\beta$ changes from zero to $-1$.  Fixing all parameters to our best fit cosmology except  $\xi_c$ or the absolute value of $\beta$  will have almost the same kind of impact in the matter power spectrum.   The growth  of density perturbations are significantly affected  by the variation of   $\beta$  or $\xi_{c}$. This is because the amount of dark matter is bigger than the fraction of dark energy at early times  when $\xi_{c}>0$ and $\beta<0$. As a result the epoch  of  matter-radiation equivalence happens earlier than the standard case so  only the very small scale perturbations on scale $k>k_{eq}$ have time to enter the horizon and grow during the radiation domination. Hence, the most obvious effect is that  the turnover in the power spectrum (which is associated with the scale  that entered the horizon when the Universe became matter dominated)  is placed  on smaller scales. In addition,  the amplitude of $P(k)$ is considerably affected   as can be noticed from  Figs. (\ref{fig:p1})- (\ref{fig:p2}). We must stress that baryons also leave their impact on  the matter power spectrum due to their coupling with the photons before recombination. Indeed, they produce  a series  of  wiggles in the power spectrum [see Figs. (\ref{fig:p1})- (\ref{fig:p2})], which are related with the existence of  a well defined peak in the correlation function of galaxies, $\xi (r)$,  placed at 150 Mpc \cite{ref:obs}.  


\section{summary}

To avoid  large-scale instability at early times  associated with  the growth of dark energy perturbations, we have employed the interacting PPF method for examining a new kind of  interaction which includes spatial variation of Hubble  function along with spatial variation of scale factor. Interestingly enough,   we found   that the transfer of energy-momentum can flow from dark energy to dark matter only, allowing to preserve dark energy and dark matter densities as positive quantities for all times. Concerning  matter perturbations,  we have shown that the overall growth rate of matter is strongly affected  in three different ways. Firstly, the dark matter density and the  Hubble rate at background level both deviate from the concordance model. Secondly,  an extra-term related with friction in  (\ref{eqn:DMP}) reveals that the local Hubble function is modified, and  thirdly  the gravitational constant is amended also. The aforesaid effects are essentials for understanding  the cosmological constraint based on  the overall  growth rate function $f_m=d{\rm ln~ \delta_{\rm m}} /d{\rm ln~ a}$  provided is connected with the measurements of  redshift space distortions through a bias-independent  quantity called $f(z)\sigma_{8}(z)$ \cite{ref:dis3}. 

Regarding  the observational side,  we performed a global fitting based on geometric and dynamical probes by combining ${\rm Planck 2013+WMAP9+JLA+BAO+RSD+HST}$ data [see Table (\ref{tab:results})]. The joint analysis  showed  that a phantom equation of state is preferred at 1 $\sigma$ level and the amount of dark matter differs from the Planck 2013 estimation in $0.11\%$ while the Planck 2015 doubles this difference after  the Planck EE+lowP  data  is used into the analysis \cite{ref:planck15}. Such disagreement can be easily understood by taking into account that the   interacting mechanism described here only allows the exchange of energy from dark energy to dark matter.  Another important ingredient  concerns to  the particular kind of power law form  selected by the data  and the strength of coupling constant. More precisely,  the joint analysis of ${\rm Planck 2013+WMAP9+JLA+BAO+RSD+HST}$ data [see Table (\ref{tab:results}) and Fig. (\ref{fig:pb})] showed that a large coupling constant cannot be ruled out in our model provided,  $\xi_c  =  0.34403427_{-    0.18907353}^{+    0.14430125}$ at 1 $\sigma$ level. At this point, we must note that  previous estimations give a coupling constant near ${\cal O}(10^{-3})$  \cite{ref:xinz}, \cite{ref:me}, even though the perturbed expansion Hubble rate term is included into the analysis  \cite{ref:yangf}.  All the combined data (${\rm Planck 2013+WMAP9+JLA+BAO+RSD+HST}$) singles out  a  power law form  with negative exponent, $\beta=   -0.50863232_{-    0.40923857}^{+    0.48424166}$ at 1 $\sigma$ level, pinpointing the  crucial effect of the interaction at early times. 

We  have examined how the interaction affects the temperature CMB spectrum profile, focusing in  the position of acoustic peaks and their  amplitudes. A similar analysis was carried out for the matter power spectrum  where the target   was to explore the position of turnover scale. Varying only the coupling constant, we found that the position of the first peak in the TT power spectrum is shifted because the fraction of dark matter increases linearly with $\xi_{c}$ [ see Fig (\ref{fig:c1})]. As the $\beta$-exponent varies only  a similar effect can be pinpointed [see Fig (\ref{fig:c2})]. Besides, the amplitude of the TT CMB spectrum at low multiples is sensitive to the value taken by the coupling constant, indicating that  the integrated Sachs-Wolfe must be included in a near future analysis \cite{ref:dis7}.  We also found an interesting effect in the matter power spectrum by  varying $\xi_{c}$ or $\beta$. It turned out that   the scale of matter and radiation equality is shifted further from the present era when the amount of dark matter is increased by  varying $\xi_{c}$ or $\beta$, and therefore a shift of the turnover point toward to smaller scale is detected [see  Figs. (\ref{fig:p1})- (\ref{fig:p2})].

\vspace{1.cm}
\acknowledgments
L.X is supported in part by NSFC under the Grants No. 11275035 and ``the Fundamental 
Research Funds for the Central Universities'' under the Grants No. DUT13LK01. 
M.G.R is partially supported by CONICET.
We acknowledge the use of the  \textsf{CosmoMC} and \textsf{CAMB}  packages \cite{ref:code2}, \cite{ref:code1}.
We acknowledge the use of \textsf{CCC} for performing the statistical analysis.
\appendix


\end{document}